\documentclass[12pt]{article}
\usepackage{graphicx}
\def\dis{distribution}
\def\pt{$p_T$}
\def\eq{Eq.\ (\ref}

\def\ed{$\eta_\Delta$}
\begin{document} 
\begin{center}  {\Large {\bf A Parton-Based Description of Forward-Backward\\ Correlation in $pp$ Collisions}}
\vskip .75cm
 {\bf Rudolph C. Hwa$^1$ and C.\ B.\ Yang$^{1,2}$}
\vskip.5cm
{$^1$Institute of Theoretical Science and Department of
Physics\\ University of Oregon, Eugene, OR 97403-5203, USA\\
\bigskip
$^2$Institute of Particle Physics, Hua-Zhong Normal
University, Wuhan 430079, P.\ R.\ China}
\end{center}\vskip.5cm
\begin{abstract} 
Forward-backward correlation in $pp$ collisions is studied in an approach that emphasizes the partonic scattering angles and circumvents the intractable problem related to the transverse momenta that are low. Assuming the back-to-back scattering of partons to be the origin of hadronic correlation, the properties of forward-backward multiplicity covariance can be derived essentially independent of details of hadronization. The range of correlation in pseudo-rapidity emerges from the study without any dynamical input, thus dispelling the notion that correlation length has any fundamental significance. An attempt is made to relate the results to the two-component structure seen in autocorrelation.
\end{abstract}

\section{Introduction}
Correlations among particles produced at high energies have always been a subject of great interest, starting from the beginning in hadronic collisions \cite{kd}, and more recently in nuclear collisions \cite{cf}. Over the years an abundant supply of experimental data have been accumulated, some of which have very high statistics, especially in the work done at relativistic heavy-ion collider (RHIC). By comparison, theoretical studies of the correlation phenomena in the bulk (excluding high-\pt\ jets) have been meager. Before the basic processes in hadron collisions mainly at low \pt\ are fully understood in a way that can be accepted as having a solid theoretical foundation, the profusion of nuclear data has inundated the subject that has no clear theoretical guidance, resulting in conflicting interpretations at times. The terms ``long-range" and ``short-range" correlations have been used, primarily in reference to what have been observed, rather than as properties of correlations understood at the level of parton interaction.

STAR collaboration has produced extensive and detailed data on correlations   in nuclear collisions \cite{jd1, jd2, jd3, tat, jd4} as well as in $pp$ collisions \cite{pt, pt1, pt2}. The analyses have been done on autocorrelations in the angular difference variables $\eta_\Delta=\eta_1-\eta_2$ and $\phi_\Delta=\phi_1-\phi_2$. The method used is to invert the scale-dependent $\left<p_T\right>$ fluctuations \cite{tpp, tp1, tp2}, yielding a rich structure in correlations not seen by any other method. Theoretical interpretation of the result is unfortunately grossly out of step with the growth of experimental information. Earlier models, such as the dual parton model \cite{dpm}, were adequate to treat the problem at a level commensurate with the coarseness of the data available at the time, but no recent attempt has been made to calculate more accurately the basic correlations in $pp$ collisions and to compare with the latest data. Soft production is still based on the idea of string fragmentation, the implementation of which is not much different from the original Lund model \cite{lund}. On nuclear collisions efforts have been made to incorporate string fusion \cite{aap, aps}, but no explicit calculation on correlation has been done. Very recently, the possibility of forward-backward correlation in the framework of color glass condensate has been advanced \cite{amp} , but without quantitative result yet to be compared with data.

In view of the present theoretical status described above, a  model-independent calculation on any portion of the problem would lead to an improvement of that status. There is, however, no feasible formalism for calculating soft QCD processes involving multiparticle production at low \pt. Instead of starting from first principles, it is sensible to examine relevant hints from the data and restrict the scope to the issues most salient for correlations. It is furthermore reasonable to start with $pp$ collisions before tackling the complexity associated with nuclear collisions. Careful analysis of the high-statistics data on the inclusive \pt\ \dis\ in $pp$ collisions at $\sqrt s=200$ GeV has revealed a two-component picture of the particle production process essentially independent of theoretical models \cite{ja5}. Inferences on the properties of parton scattering are made. The dominant soft component gives rise to a correlation behavior in \ed\ that is drastically different from the contribution from the minor, but significant, harder component, as revealed by the Porter-Trainor analysis of autocorrelation for $pp$ collisions \cite{pt, pt1, pt2}. Although the autocorrelation measure is the Pearson's correlation coefficient that is the normalized version of the covariance of multiplicity densities, the \ed\ dependence is primarily the same as the forward-backward multiplicity covariance $D_{fb}^2$ that has been determined without averaging over the rapidity sum $\eta_1+\eta_2$ \cite{sr}. 

We shall focus on just that piece of data on $D_{fb}^2$ and relate it to the basic partonic scattering process that has back-to-back correlation. The connection between scattering angle and pseudo-rapidity is carefully done at the parton level to reveal the basic structure in the \ed\ dependence after hadronization is incorporated to relate the  partonic properties  to the hadronic observables. Since the processes are at low \pt, there are inevitably some free parameters that are not fixed by our current understanding of soft physics. While the strength of correlation is adjustable to fit the data, there are no adjustable parameters to specify the range of correlation. Thus an understanding of the \ed\ dependence will be the main finding in this study that is essentially independent of models. Hadronization will be considered in the context of recombination, but details are not important in any essential way. A significant outcome of our study is the recognition that parton and hadron pseudo-rapidities can be sufficiently different that partons in one region may give rise to hadrons in another region, thereby cross-fertilizing forward and backward windows that are not too far apart. A correspondence of our result with the two components in the Porter-Trainor analysis will be attempted at the end of our study.

\section{Two-parton joint distribution}

We start with a brief discussion of the kinematical relationship between the parton and hadron variables.
Since the forward-backward correlation (FBC) is measured in terms of pseudo-rapidity without reference to the transverse momenta \pt\ of the detected hadrons, it is clear that the correlation properties are dominated by the hadrons produced at low \pt, since high-\pt\ particles are severely suppressed. Since pQCD is not valid at low \pt, there is no reliable theoretical formalism in which rigorous calculation can be made. Nevertheless, quarks and gluons are the constituents that interact, and those partons before and after scattering are the intermediary that bridges the initial state of the nucleons and the final state of multiparticles. We therefore seek some simple description of the parton state before hadronization in the hope that the observed FBC can be related to some basic properties of the partons. Since the hadronic \pt\ is not specified in the measurement of FBC, there is no point in emphasizing the partonic transverse momenta $k_T$, especially when their behavior is not calculable. Hence, our focus will be on the angular properties of the partons and hadrons.

Let the parton and hadron momenta be denoted by $\vec k$ and $\vec p$, respectively, and let their corresponding polar angles be $\theta$ and $\Theta$. If the angle between $\vec k$ and $\vec p$ is denoted by $\psi$, then we have
  \begin{eqnarray}
\cos\Theta=\cos\theta\cos\psi+\sin\theta\sin\psi\cos\varphi,   \label{1}
\end{eqnarray}
where $\varphi$ is the solid angle between the plane containing  $\vec k$ and $\vec p$ and the plane containing  $\vec k$ and the $z$ axis. The $\psi$ dependence in the problem is a property of the hadronization process. At low \pt\ parton recombination is far more efficient, thus dominant, compared to fragmentation, which requires higher $k_T$ partons. The dynamical mechanism need not be specified at this stage of geometrical consideration, except to mention that another parton beside the one at $\vec k$, not necessarily collinear, has to be picked up to form the hadron at $\vec p$. Let the hadronization cone be described by a Gaussian \dis\ in $\psi$
\begin{eqnarray}
g(\psi)=g_0\exp(-{\psi^2\over 2\sigma^2}), \label{2}
\end{eqnarray}
where the normalization factor $g_0$ is determined by
\begin{eqnarray}
\int_0^{2\pi}d\varphi\int_0^{\pi/4}d\psi \sin\psi\ g(\psi)=\left<N\right>.  \label{3}
 \end{eqnarray}
$\left<N\right>$ is the average number of hadrons produced by a parton. The cone width  $\sigma$ is expected to be small for $pp$ collisions, and larger for AA collisions. For $\sigma$ much smaller than the range of angular integration in Eq.\ (\ref{3}),  we approximate $\sin \psi$ by $\psi$ so the integration can be readily carried out, yielding
 \begin{eqnarray}
g_0={\left<N\right>\over 2\pi\sigma^2}. \label{4}
\end{eqnarray}
The differential form of Eq.\ (\ref{3}) is
\begin{eqnarray}
{dN\over d\varphi d\cos\psi}=g(\psi). \label{5}
\end{eqnarray}
At this stage of our discussion of hadronization it really does not matter whether hadrons are formed by recombination or fragmentation, since a hadronization cone is relevant in both cases.
We postpone further discussion about hadronization until the following section, since what is described above is sufficient to lead us from the hadrons to the partons. 

We now consider the two-parton \dis\ of back-to-back scattering at low $k_T$ in the CM system of the partons. Starting from the simplest possible form at this point, with allowance  for complications to be included later, we write
\begin{eqnarray}
{dn\over dz_1dz_2}=A_0\ \delta(z_1+z_2), \label{6}
\end{eqnarray}
where $z_i=\cos\theta_i$, assuming temporarily that the partonic rest frame is the same as the $pp$ CM system.  Since the two partons are exactly back-to-back, it is obviously a possible source of correlation at the partonic level. There can, of course,   be other types of partonic interaction not describable in a simple form as in \eq{6}). However, it is not our problem here to provide a listing of all possible expressions of those interactions.
 Our aim is to investigate to what extent the observed FBC at the hadronic level can be traced back to Eq.\ (\ref{6}).

The $\delta(z_1+z_2)$ function is, of course, to constrain $z_2=-z_1$ that is only valid if the CM systems of partons and hadrons coincide. In reality the proton has a wide \dis\ of partons at low $x$, so any pair of them from the two initial protons can have varying degree of mismatch between the two systems. Thus in the $pp$ system we must allow $\delta(z_1+z_2)$ to be broadened, but the way in which it is to be done is not calculable, since these are all low-$Q^2$ partons. We use a one-parameter description of the broadened \dis\ to replace $\delta(z_1+z_2)$:
\begin{eqnarray}
D_m(z_1+z_2)={1\over N_m}\left[1-{1\over 4}(z_1+z_2)^2\right]^m,   \label{7}
\end{eqnarray}
which is forced to vanish at the kinematic limit $|z_1+z_2|=2$ and is normalized such that 
\begin{eqnarray}
N_m=\int_{-2}^2 dz D_m(z)=2B(1/2,m+1),  \label{7.1}
\end{eqnarray}
$B(a,b)$ being the Beta function. 
This \dis\ is roughly Gaussian shaped in the finite interval that $z_1+z_2$ is allowed to vary, and can be broad when $m$ is small.  Since $D_0$ is a constant,  we know that in the limit $m\rightarrow 0$, $D_m(z_1+z_2)$ cannot give rise to  any correlation between the two partons. Thus for the correlated part, we must subtract the uncorrelated part and introduce
\begin{eqnarray}
C_m(z_1+z_2)=D_m(z_1+z_2)-D_0,  \label{7.2}
\end{eqnarray}
which is what we shall use in place of $\delta(z_1+z_2)$ for the correlation to be calculated below.

Let the pseudo-rapidity of a parton be denoted by $\zeta$, i.e.,
 \begin{eqnarray}
\zeta=-\ln\tan\theta/2 ,   \label{8}
\end{eqnarray}
from which can be derived
 \begin{eqnarray}
z=\cos\theta=\tanh\zeta,  \quad\qquad dz= \cosh^{-2}\zeta\ d\zeta. \label{9}
\end{eqnarray}
Using this in Eqs.\ (\ref{6}) and (\ref{7.2}) yields
 \begin{eqnarray}
{dn_c\over d\zeta_1d\zeta_2}={A_0\over \cosh^2\zeta_1\cosh^2\zeta_2}\ C_m(\zeta_1,\zeta_2), \label{10}
\end{eqnarray}  
   where
  \begin{eqnarray}
C_m(\zeta_1,\zeta_2)={1\over N_m}[1-{1\over 4}(\tanh\zeta_1+\tanh\zeta_2)^2]^m-{1\over 4}.
\label{11}
  \end{eqnarray}
The subscript $c$ in Eq.\ (\ref{10}) is to emphasize that it is the correlated part of the two-parton \dis\ we are describing.

\section{Two-hadron joint distribution}

Having obtained an expression for the correlated two-parton \dis\ in $\zeta_1$ and $\zeta_2$, we now develop from it the correlated part of the two-hadron \dis. Equation (\ref{1}) gives the relationship between the hadronic polar angle $\Theta$ in terms of the partonic polar angle $\theta$, with $\psi$ being the hadronization cone angle. Since soft hadrons with $p_T$$^<_{\sim} 1$ GeV/c dominate FBC, the participating partons that hadronize by recombination have transverse momenta $k_T$ in the range $^<_{\sim} 0.5$ GeV/c. For a quark with $\vec k$ that hadronizes, it picks up an antiquark with momentum $\vec k'$ and forms a pion with momentum $\vec p=\vec k+\vec k'$. The vectors $\vec k$ and $\vec k'$ need not be exactly collinear, but their angular difference cannot be too large, since the pion size is finite. In momentum space the uncertainty in the pion wave function is of the order of the pion mass, so the angular difference between $\vec k$ and $\vec k'$ is of order 0.14/0.5. The angle $\psi$ between $\vec k$ and $\vec p$ is about half that much. Thus the hadronization cone width $\sigma$ in Eq.\ (2) is roughly between 0.1 and 0.2.

The hadronization process considered here implies that only one pion is produced per inclusive parton, since we are not discussing the total number of hadrons formed from a fixed pool of partons. Thus the normalization integral in \eq{3}) is $\left<N\right>=1$. The precise value of  $\left<N\right>$ is unimportant in the situation where the parton density is not precisely known, exemplified by the undetermined parameter $A_0$  in \eq{6}). We have not  been thorough in our discussion about the recombination process, nor have we mentioned specifically the role that gluons play. The details of such processes, including the conversion of gluons to quarks before hadronization, have been studied extensively before \cite{rh,hy}. They are omitted here since such details would not contribute to the clarification of any crucial issues at hand and would only distract the flow of our main concern here.

For a fixed parton momentum $\vec k(\theta)$ the probability of finding a hadron at angles $(\psi,\varphi)$ relative to it is given by \eq{5}). To guarantee that \eq{1}) is satisfied for the hadron at $\Theta$, we define the hadron \dis\ per unit $\cos\Theta$ as 
\begin{eqnarray}
\tilde G(\theta,\Theta)={dN\over d\cos\Theta}=\int_0^{2\pi} d\varphi \int_0^{\pi/4} d\psi {dN\over d\varphi d\psi}\delta(\cos\Theta-\cos\theta\cos\psi-\sin\theta\sin\psi\cos\varphi) \label{12}
\end{eqnarray}
such that
\begin{eqnarray}
\int d\cos\Theta \tilde G(\theta,\Theta)=1,  \label{13}
\end{eqnarray}
where the range of integration corresponds to the angles in \eq{3}). Carrying out the integration over $\varphi$ in Eq.\ (\ref{12}) we obtain
\begin{eqnarray}
\tilde G(\theta,\Theta)=2\int d\psi \sin\psi\,g(\psi)|\sin^2\theta\sin^2\psi-(\cos\Theta-\cos\theta\cos\psi)^2|^{-1/2}.  \label{14}
\end{eqnarray}
 In the realistic 3D geometry of the experiments the angular measure is in $d\cos\Theta\,d\phi$. The azimuthal angle $\phi$ around the beam axis is integrated over $2\pi$ in both \cite{pt2} and \cite{sr}, and is of no concern here. 
 
 With the definition $Z=\cos\Theta$ the hadron \dis\ in $Z$ is related to the parton \dis\ in $z$ as
  \begin{eqnarray}
{dN\over dZ}=\int dz {dn\over dz} \tilde G(z,Z).  \label{15}
\end{eqnarray}
In terms of the hadron pseudo-rapidity $\eta$, as in \eq{9}),
\begin{eqnarray}
Z=\tanh\eta, \qquad\quad  dZ=\cosh^{-2}\eta\ d\eta,   \label{16}
\end{eqnarray}
we have
 \begin{eqnarray}
{dN\over d\eta}=\int d\zeta {dn\over d\zeta} {1\over \cosh^2\eta}\tilde G(z(\zeta),Z(\eta)).  \label{17}
\end{eqnarray}
Defining
\begin{eqnarray}
 G(\zeta,\eta)={1\over \cosh^2\eta}\tilde G(z(\zeta),Z(\eta)),   \label{18}
\end{eqnarray}
we then have for the two-hadron \dis\
\begin{eqnarray}
{dN\over d\eta_1d\eta_2}=\int d\zeta_1d\zeta_2{dn\over d\zeta_1d\zeta_2} G(\zeta_1,\eta_1)G(\zeta_2,\eta_2). \label{19}
\end{eqnarray}
 The hadronization function $G(\zeta,\eta)$ has no simple analytical form and is non-trivial. For small $\sigma$, $G(\zeta,\eta)$ may be shaped roughly as a Gaussian, depending on $\zeta$. Because of the non-linear relationship between polar angle and pseudo-rapidity, what is symmetric in $\psi$ as in \eq{2}) cannot be symmetric in $\eta$ around $\zeta$. In Fig.\ 1 we show some illustrative examples of $G(\zeta,\eta)$ for $\zeta=0.2$ and $\sigma=0.1, 0.2$ and 0.5. As $\sigma$ increases, the asymmetry in $\eta-\zeta$ develops in the wings, and becomes more significant for higher $\zeta$ (not shown). The physics of hadronization is in the physical 3D momentum space involving $\psi$, not in $\zeta$ or $\eta$. In AA collisions $\sigma$ can be large and $G(\zeta,\eta)$ can be wide in $\eta-\zeta$. It is then unreliable to assume that a window in $\eta$ for the detected hadrons corresponds to a similar window in $\zeta$ for the originating partons.
  
 Returning to the two-hadron \dis\ given in \eq{19}) we note that the two partons at $\zeta_1$ and $\zeta_2$ hadronize independently, but because of the finite width in $\eta-\zeta$ of the $G$ functions, it is possible that for any given window in $\eta$ the detected hadrons in any given event can originate from both partons separately at different $\zeta_1$ and $\zeta_2$. In other words, hadronization can ``spill over" from positive to negative sides, and vice-versa, even if the forward and backward windows are located symmetrically on the two sides of $\eta=0$.
 
 Putting \eq{10}) in (\ref{19}) we obtain for the correlated two-hadron \dis\
    \begin{eqnarray}
{dN_c\over d\eta_1d\eta_2}=\int d\zeta_1d\zeta_2{A_0\over \cosh^2\zeta_1\cosh^2\zeta_2}C_m(\zeta_1,\zeta_2) G(\zeta_1,\eta_1)G(\zeta_2,\eta_2).  \label{20}
\end{eqnarray}
Note that without the $C_m(\zeta_1,\zeta_2)$ function the integral is factorizable, resulting in no correlation. Thus the non-factorizable $C_m(\zeta_1,\zeta_2)$  is the source of FBC. The inverse-square factor $(\cosh\zeta_1 \cosh\zeta_2)^{-2}$ suppresses the large $\zeta$ contribution, so $dN_c/d\eta_1d\eta_2$ appears to have no long-range correlation. However, that factor is the Jacobian of the transformation from $z$ to $\zeta$ in \eq{9}) and has no dynamical content. Thus the designation of such terms as short- or long-range correlation can be misleading.

\section{Forward-backward correlation}

To get FBC we integrate \eq{20}) over $\eta_1, \eta_2$ with $\eta_1$ in the forward window, and $\eta_2$ in the backward window, defined to have widths $\delta\eta$ and spaced symmetrically apart from $\eta=0$ with $\eta_\Delta$ being the distance between the centers of the windows. Thus the correlated part of the forward and backward multiplicities is
 \begin{eqnarray}
\left< N_F N_B\right>_c(\eta_\Delta)=\int_{\eta_-}^{\eta_+}d\eta_1\int_{-\eta_+}^{-\eta_-}d\eta_2 {dN_c\over d\eta_1d\eta_2} \hspace{6cm} \nonumber \\
= \int d\zeta_1\int d\zeta_2 {A_0\over \cosh^2\zeta_1\cosh^2\zeta_2}C_m(\zeta_1,\zeta_2)H_F(\zeta_1,\eta_\Delta) H_B(\zeta_2,\eta_\Delta),  \label{21}
\end{eqnarray}
where $\eta_-=(\eta_\Delta-\delta\eta)/2$ and $\eta_+=(\eta_\Delta+\delta\eta)/2$ , and
  \begin{eqnarray}
H_F(\zeta_1,\eta_\Delta)=\int_{\eta_-}^{\eta_+}d\eta_1  G(\zeta_1,\eta_1) ,  \label{22} \\
 H_B(\zeta_2,\eta_\Delta)= \int_{-\eta_+}^{-\eta_-}d\eta_2 G(\zeta_2,\eta_2).    \label{23}
\end{eqnarray}
      It is clear in \eq{21}) that hadronization is described entirely by the $H_F$ and $H_B$ functions.  The hadronization of the two partons at $\zeta_1$ and $\zeta_2$ can allow a forward (backward) parton into a backward (forward) window when $\eta_\Delta$ is small, since $H_F$ and $H_B$ are not narrow functions.
      
	To illuminate the properties of hadronization, we show in Fig.\ 2 $H_F(\zeta,\eta_\Delta)$ vs $\zeta$ for fixed $\eta_\Delta=0.4$ and $\delta\eta=0.2$, and for three values of $\sigma$. 
All curves peak at $\zeta=\eta_\Delta/2=0.2$, which is the location of the window extending from 0.1 to 0.3, as indicated by the shaded interval. The solid curve shows that even for the narrow cone width of $\sigma=0.1$ partons significantly outside that window can contribute.
It is even wider for the dashed curve.  Although large $\sigma$ (such as 0.5) is not relevant to $pp$ collisions, it shows that the value of $\zeta$ of the contributing parton can be greater than 1, i.e., outside the detector coverage of STAR.

	The covariance of $N_FN_B$ that is measured is denoted by $D_{fb}^2$: 
	\begin{eqnarray}
D_{fb}^2=\left<N_FN_B\right>-\left<N_F\right>\left<N_B\right>, \label{24}
\end{eqnarray}
whose $\eta_\Delta$ dependence in the data \cite{sr} is shown in two figures below. Identifying $D_{fb}^2=\left<N_FN_B\right>_c$ with \eq{21}), we can fit the  data by varying the relevant parameters in the problem. Before discussing how that is done, it is  important to note first that the decrease of $D_{fb}^2$ with increasing $\eta_\Delta$ is strongly affected by  the $(\cosh\zeta_1\cosh\zeta_2)^{-2}$ factor in the integrand in \eq{21}). Changing that factor artificially to a weaker power diminishes the rate of decrease of $\left<N_FN_B\right>_c$ with $\eta_\Delta$ and will not fit the data. Since there is nothing to adjust in that factor to reflect the nature of interaction, the range of correlation that one may naively infer from the data has no dynamical meaning. 

There are essentially only two parameters to adjust in the problem: $A_0\ {\rm and} \ m$. $A_0$ describes the strength of  parton interaction at low \pt,   and $m$ specifies the spread of the parton rest frame relative to the $pp$ CM system, thereby affecting the observed properties of correlation. $\sigma$ is between 0.1 and 0.2, and will not affect the final result  sensitively. We adopt the following fitting strategy. First,  we fix $A_0=1,\ \sigma=0.2$, and vary $m$ to see the corresponding dependence of $\left<N_FN_B\right>_c(m, \eta_\Delta)$ on \ed. That is shown in Fig.\ 3.  Evidently, there is strong dependence on $m$. For narrower $C_m(\zeta_1,\zeta_2)$ at higher $m$, the hadronic FBC is stronger, as it should for fixed $A_0$.
However, the dependence on \ed\ appears to be universal. We plot the ratios 
\begin{eqnarray}
R_m(\eta_\Delta)={\left<N_FN_B\right>_c(m)\over \left<N_FN_B\right>_c(m=1)}    \label{25}
\end{eqnarray}
in the inset and find that they are essentially independent of \ed, except when $m=10$ or larger. We do not expect $D_m$ to be narrow, so $m$ should not be large. For smaller $m$, $R_m$ increases approximately linearly with $m$. A fit of the whole range of $m$ studied results in
\begin{eqnarray}
R_m=0.0517+1.0335m-0.1083m^2+0.005m^3,  \label{25.1}
\end{eqnarray}
at $\eta_\Delta=0.6$.
It is clear then to fit the data on $D_{fb}^2$ the value of $A_0$ must decrease with $m$ as $R_m^{-1}$. Let us then define 
\begin{eqnarray}
B_m\equiv A_0R_m(\eta_\Delta=0.6),  \label{26}
\end{eqnarray}
which we expect to be a rather stable variable to use to fit $D_{fb}^2$ for a wide range of $m$.
Still holding $\sigma=0.2$, we obtain the results shown collectively by the solid line in Fig.\ 4 for $m=0.5, 1, 2$. There is no dependence on $m$. The value of $B_m$ is 
\begin{eqnarray}
B_m=14, \qquad\qquad m<3.   \label{27}
\end{eqnarray}
The reproduction of the data in their \ed\ dependence is very good, except for the highest point at \ed=1.6. We emphasize that we only vary $B_m$ to fit the normalization; the \ed\ dependence follows from \eq{21}) without adjustment. For $\sigma=0.1$ we fix all other parameters already considered, and obtain the dashed line in Fig.\ 4 for $m=1$. For other values of $m$ there are some small variations because $R_m$ that is used is given in \eq{25.1}), which was calculated for $\sigma=0.2$. But the difference is miniscule ($<5\%$).
The dependence on $\sigma$ is evidently not strong, so we shall hereafter adhere to the value $\sigma=0.2$.

To summarize our  result so far, we see that $A_0$ and $m$ are strongly correlated in the fitting procedure. $A_0C_m(\zeta_1,\zeta_2)$ is the source of hadronic correlation that replaces the $A_0\delta(z_1+z_2)$ in \eq{6}).  The  smaller $m$ is, the larger must be  $A_0$ to compensate for the spread of   $C_m(\zeta_1,\zeta_2)$  in order to  account for the observed magnitude of FBC. We have  combined the two  into one parameter $B_m$ and succeeded in fitting the data. But the adjustment is only in the magnitude of the correlation. The important point is that the basic dependence of $\left<N_FN_B\right>_c$ on \ed\ is due to the  factor  $(\cosh\zeta_1\cosh\zeta_2)^{-2}$ in \eq{21}) that is not adjustable and is the origin of the universality seen in Fig.\ 3.

Let us, for definiteness, consider the case $m=1$, for which $R_1=1$, so  $A_0=14$. Using this number in \eq{6}) and integrate it over $-1\le(z_1,z_2)\le1$, we obtain 28 correlated pairs of partons. This is not a large number compared to the total number of pairs of partons $\left<n(n-1)\right>$, where $\left<n\right>$ is roughly twice the total average charged event multiplicity $\left<N\right>$ of hadrons, which is about 20. Thus only a small fraction ($<2\%$) of all parton pairs are correlated.

Returning now to \eq{6}) that expresses the two-parton \dis\ assumed from the start, one may regard that as the $s$-wave contribution to the scattering process. It is of interest to ask whether the data on FBC can admit a $p$-wave contribution as well. To that end we extend \eq{6}) to include a second term
\begin{eqnarray}
{dn\over dz_1dz_2}=[A_0+A_1(z_1^2+z_2^2)] \delta(z_1+z_2), \label{28}
\end{eqnarray}
which is symmetric under $z_1\leftrightarrow \pm z_2$.
Since all quantities considered here and in the experiment are integrated over the $2\pi$ range of the azimuthal angle $\phi$, the cross term between $s$ and $p$ waves vanishes by azimuthal symmetry. Applying \eq{28}) to \eq{21}) we replace the numerator by $A_0+A_1(\tanh\zeta_1^2+\tanh\zeta_2^2)$. The $p$-wave contribution increases with $\zeta_i$ because of the factor $(\tanh\zeta_1^2+\tanh\zeta_2^2)$, so it may lift the upper end of the \ed\ \dis\ of $\left<N_FN_B\right>_c$ for a possibly better fit of the data. For values of $A_0$ and $A_1$ to be specified below, the  contributions of the $s$- and $p$-wave components are   shown by the dashed and  dashed-dotted lines, respectively, in Fig.\ 5; the latter indeed increases with \ed. 
The solid line represents their sum, whose fit of the data is about comparable to that in Fig.\ 4.
However, we have another motive for considering the additional $p$-wave contribution.

To explain that, it is best to rewrite \eq{28}) first as 
\begin{eqnarray}
{dn\over dz_1dz_2}=[A_s+A_h(z_1,z_2)]\delta(z_1+z_2),   \label{32}
\end{eqnarray}
where
\begin{eqnarray}
A_s&=&A_0-0.4\,A_1 ,  \label{33}  \\
A_h(z_1,z_2)&=&A_1(0.4+z_1^2+z_2^2) .  \label{34}
\end{eqnarray}
The inclusion of $0.4A_1$ in \eq{34}) is for the purpose of rendering the $A_h$ contribution roughly constant in \ed. That is shown by the dotted line in Fig.\ 5. The overall fit of the data is now achieved by choosing 
\begin{eqnarray}
A_0=13.2, \qquad\qquad A_1=0.2 A_0=2.64.   \label{35}
\end{eqnarray}
The dotted line is the contribution from $A_h(\zeta_1,\zeta_2)$ magnified by a factor of 2 for visual clarity, as is done for the dashed-dotted line.  Without the multiplier, $A_h$ is roughly $A_s/10$ at small \ed; that is the constraint that leads to the determination of $A_1$ relative to $A_0$ in \eq{35}).  Although the fit of the data on $D_{fb}^2$ in Fig.\ 5 is not significantly improved, we have demonstrated that the data  can be understood as a combination of two components, one of which is nearly flat in \ed.

The reason for doing the above decomposition is to establish a connection with the picture formed in the data analyzed in Ref.\ \cite{ja5}  that there exist  a soft and a hard component;  the former is larger and has a Gaussian-like decrease  in \ed, while the latter is smaller  and   roughly constant in \ed\ \cite{pt,pt1,pt2}. Our notations $A_s$ and $A_h$ are chosen to make a symbolic correspondence to those soft and hard components. The magnitude of the hard component depends on the charge multiplicity of the event class analyzed. The ratio of hard to soft components can be as much as 10\% but can be much lower.  The choice of $A_1$ in \eq{35}) and the solid line in Fig.\ 5 correspond to the maxium value that $A_1$ can have. For weaker hard components, $A_1$ is lower and the height of the dotted line is also lower accordingly. The quality of the overall fit of the data on $D_{fb}^2$ in Fig.\ 5 is largely unaffected.

The soft and hard components  in Ref.\ \cite{ja5} are determined by detailed analysis of the \pt\ behavior of the produced hadrons, the hard having $\left<p_T\right>$ only around 1.2 GeV/c, but still higher than the soft component. Our study here bypasses the issue about the transverse momenta of the partons and hadrons, which are not tractable at such low \pt, but focuses on the angular variables that are more intimately related to pseudo-rapidity. We find that the properties found in Refs.\ \cite{pt,pt1,pt2,ja5}   can be accommodated by a combination of $s$- and $p$-wave components to give the $A_s$  and $A_h$  components. Since we have had no need to question the $k_T$ aspect of the partons in the present treatment, we have no basis to judge what is soft or hard. However, it does seem reasonable that the soft component consists entirely of the $s$-wave, while the hard component includes also the $p$-wave part of the scattering.

\section{Conclusion}

We have considered hadronic FBC by relating it to the simplest form of partonic interaction that has back-to-back correlation. It is found that it gives a fairly good description of the dependence of FBC on window separation \ed\ without any adjustment of the range of correlation. The magnitude of the correlation function is fitted by varying the number of correlated pairs of partons, which turns out to be less than 2\% of all possible pairs that hadronize. The so-called ``range of correlation" in $\eta$ is mainly a consequence of the transformation from the polar angles, and has no meaning at the parton level.

Since low-\pt\ hadrons dominate any hadronic measure that does not restrict the \pt\ range of coverage, it is necessary to consider low-$k_T$ partons for which available theoretical tools are deficient. Leaving open the questions about $k_T$ and \pt, we have focused on the relationship between polar angles and pseudo-rapidities and found a meaningful way to separate different issues that include the mismatch between the partonic and hadronic CM systems in the initial state, and the hadronization of partons in the final state. Uncertain properties of those two specific issues have been investigated, and various possibilities have been considered before arriving at the final result that is mainly insensitive to those properties.

Since at $p_T<1.5$ GeV/c recombination may be more likely than fragmentation as the dominant mechanism of hadronization, we have considered the effect of non-vanishing width of the hadronization cone. We find that there is significant  cross feeding of partons from one region to hadrons in neighboring regions. To identify the partonic window   with the hadronic window is an assumption that becomes even more unreliable in nuclear collisions, where attempts have been made to infer  the nature of the bulk medium   from the observed charge fluctuation in restricted windows. 

We have attempted to make contact with the results of Porter-Trainor analysis that shows by autocorrelation   the existence of very different \ed\ behaviors for the soft and hard components at low $p_T$ $(<1.5$ GeV/c) \cite{pt, pt1, pt2}. We can identify the (soft) component having strong \ed\ dependence with our $s$-wave component and the weaker (hard) component having roughly no dependence on \ed\ with our part that  includes the $p$-wave contribution. Our major finding is that the strong \ed\ dependence does not imply short-range correlation. It is possible that the \ed-independent component may be related to multiplicity fluctuation in some way by treating the ``minijets" in the hard component at the parton level.

The parton-based approach adopted in this work is clearly different from the string model that emphasizes the role of the valence quarks in the formation of strings and the mechanism of  string fragmentation for soft production of particles. Our approach is closer to the original parton model, where low-momenta soft partons are a part of the initial state of the incident proton \cite{rf}.

With the elucidation that we have achieved for the $pp$ collision problem, it is natural to ask what light it sheds on the $AA$ collision problem. To the extent that we have investigated the subject we have found that the nuclear problem is vastly more complicated, since there
 are  many contributing factors that can influence FBC.  What we have uncovered in the $pp$ problem is likely to be overwhelmed by fluctuations related to the particle production processes  outside the realm of partonic correlation  in $pp$ collisions. It is therefore mostly a  separate problem only a part of which is connected with the topic of study here.

\section*{Acknowledgment} 
 We are grateful to Tom Trainor for extensive discussions that help us to understand the nature of autocorrelation from the analysis that he and collaborators have undertaken. We also thank Brijish Srivastava for communication on the data analysis that he and collaborators have done. This work was supported, in part,  by the
U.\ S.\ Department of Energy under Grant No. DE-FG02-92ER40972   and by National Natural Science Foundation of China under Grant No. 10475032.

\newpage
\begin{figure}[htbp]
\centering
\includegraphics[width=6in]{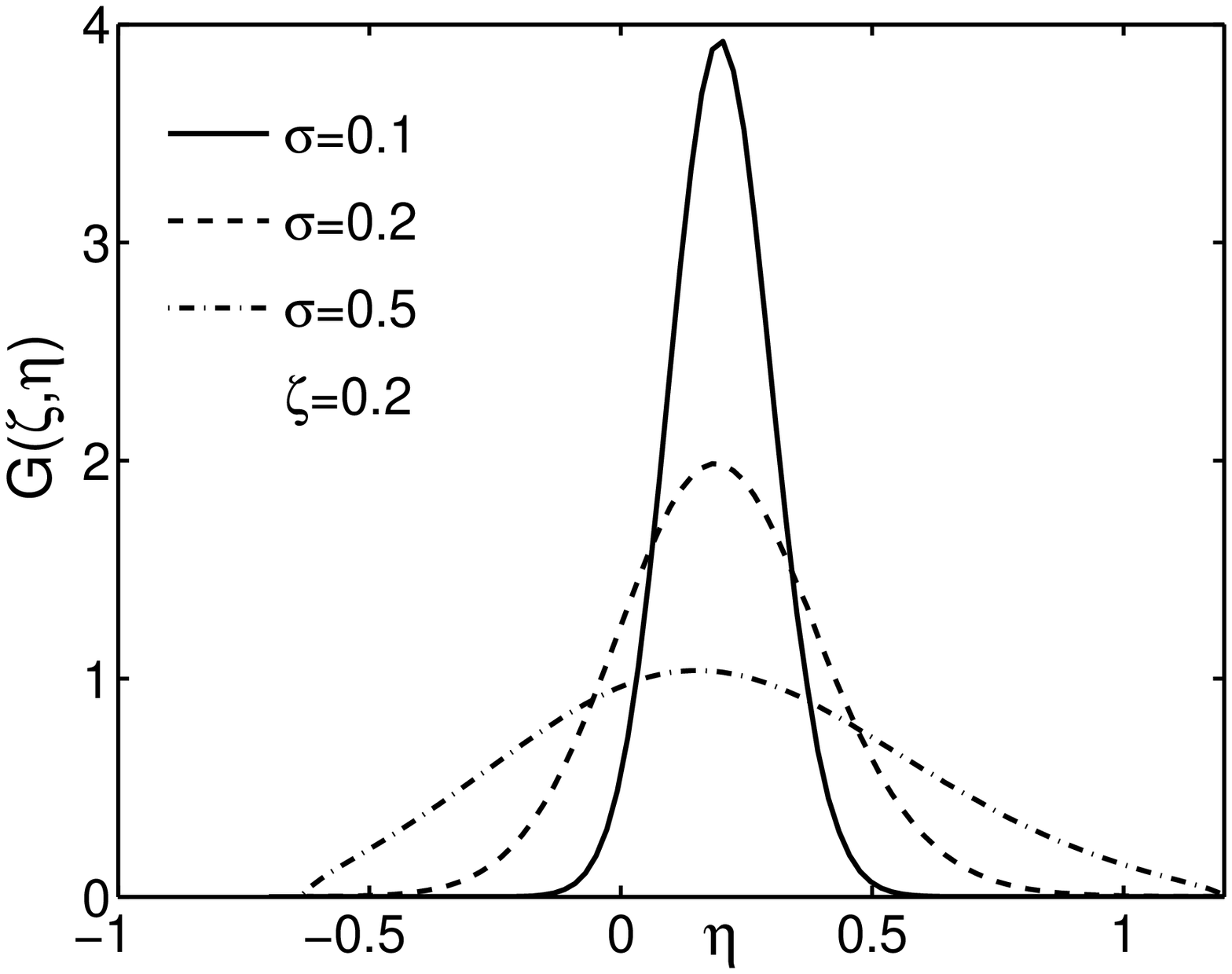}
\caption{
The \dis s of $G(\zeta,\eta)$ in $\eta$ for $\zeta=0.2$ and for three values of the cone width $\sigma$.}
\end{figure}

\newpage
\begin{figure}[htbp]
\centering
\includegraphics[width=6in]{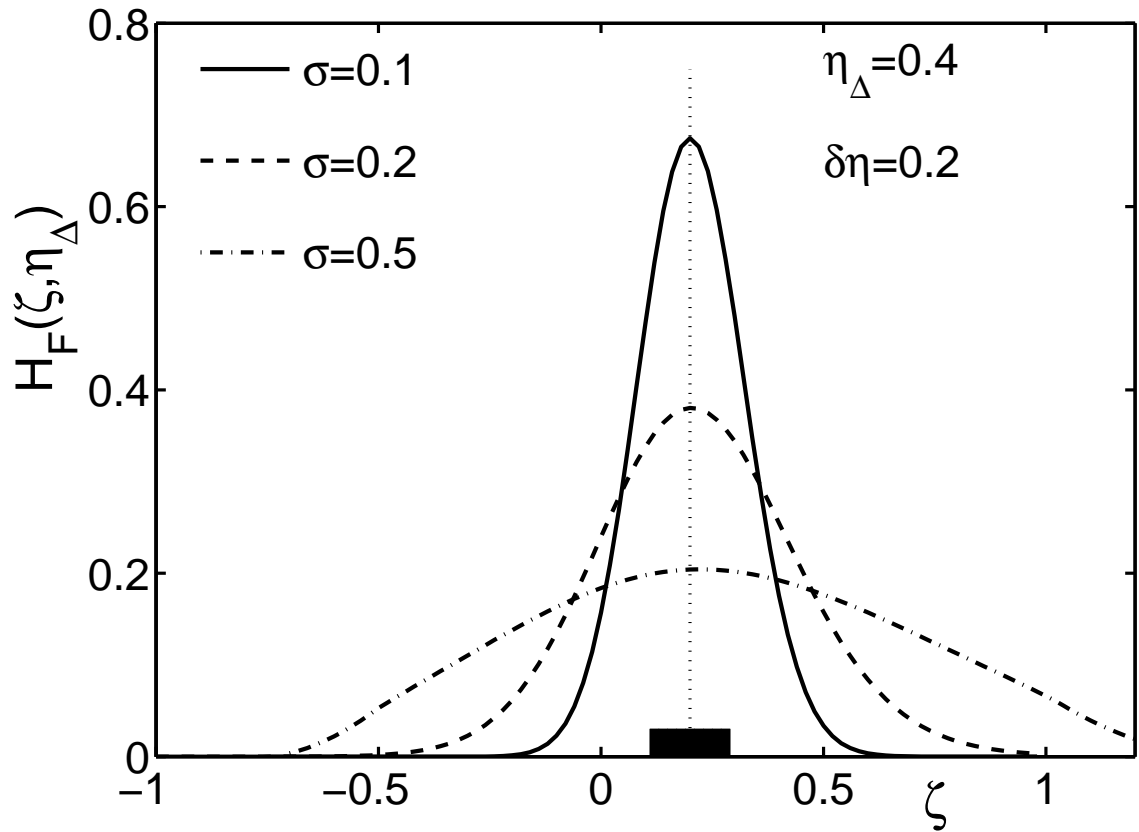}
\caption{The \dis s of $H_F(\zeta,\eta_\Delta)$ for a forward window at $\eta_\Delta/2=0.2$ and window size $\delta\eta=0.2$ indicated by the shaded interval.}
\end{figure}

\newpage
\begin{figure}[htbp]
\centering
\includegraphics[width=6in]{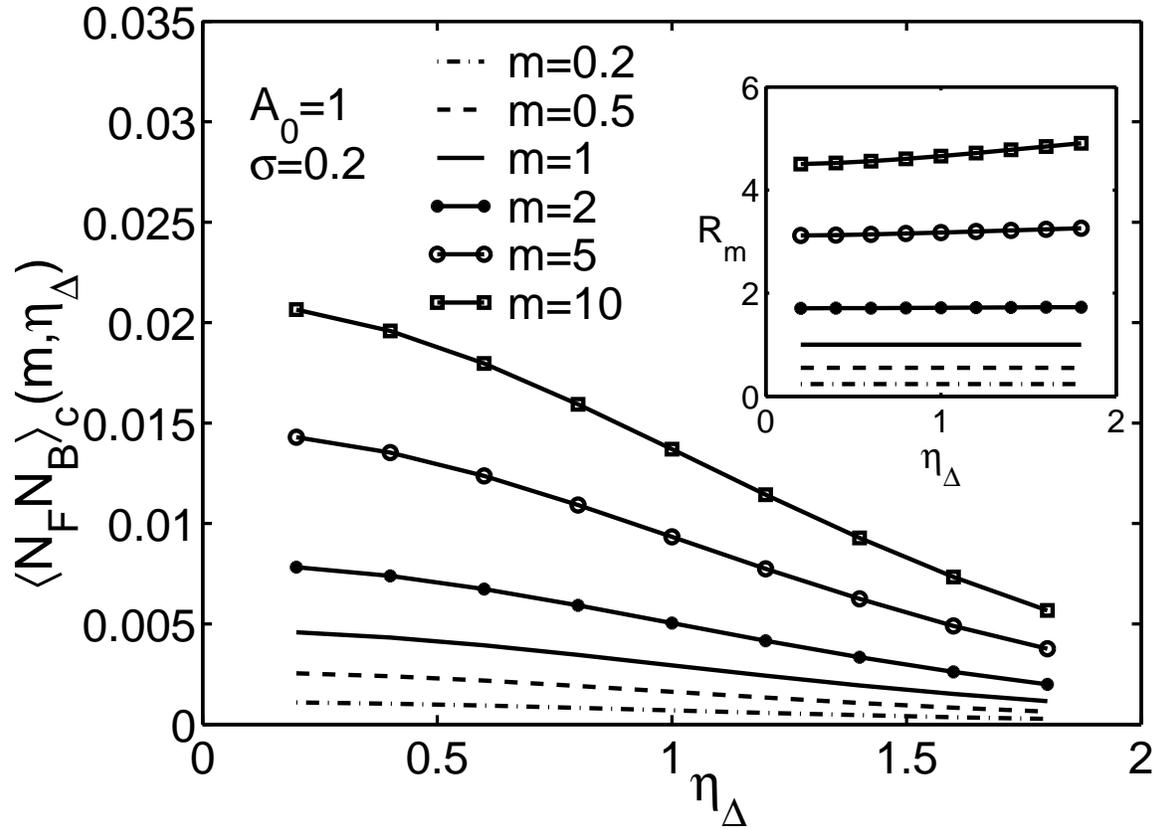}
\caption{Examples of the FBC function in \ed\ for fixed $A_0=1$ and $\sigma=0.2$, and for six values of $m$ that characterizes the width of $C_m(\zeta_1,\zeta_2)$. Ratios of those functions relative to $m=1$ are shown in the inset.}
\end{figure}

\newpage
\begin{figure}[htbp]
\centering
\includegraphics[width=6in]{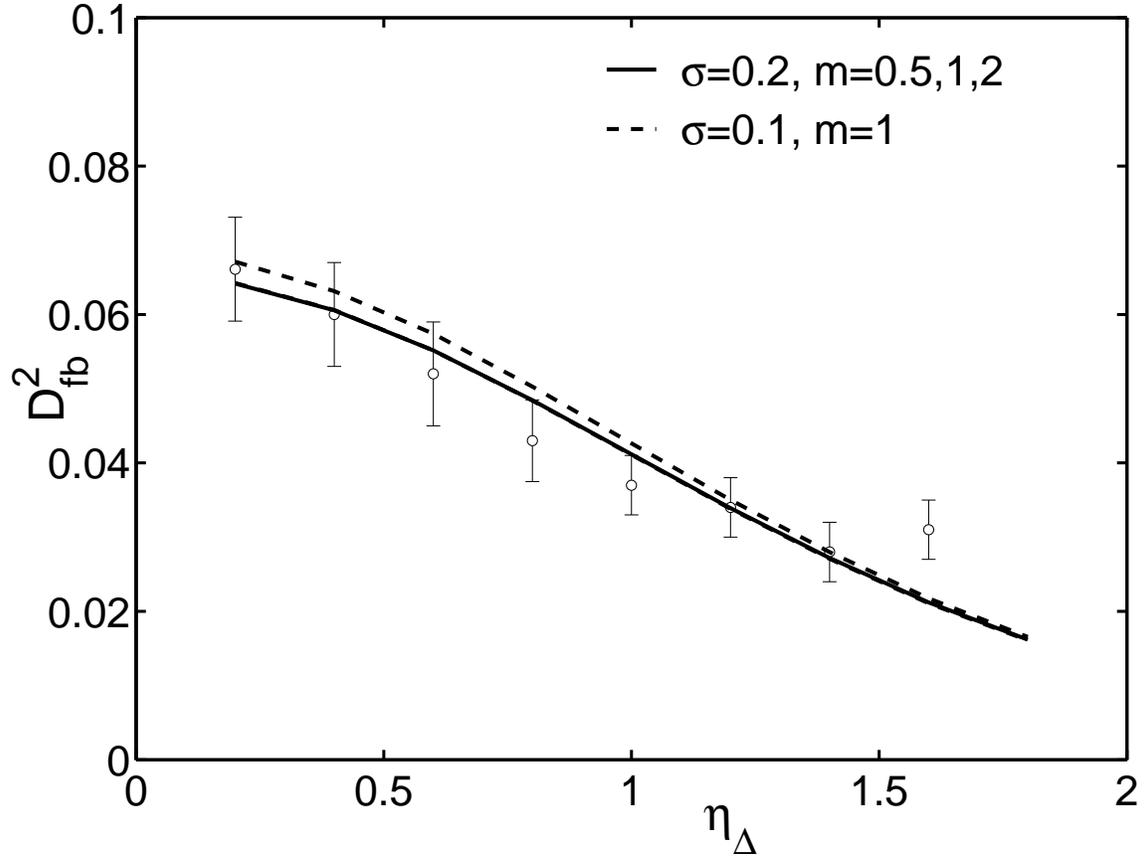}
\caption{Forward-backward multiplicity covariance versus \ed. Data are from Ref.\ \cite{sr}. The solid line shows the coalescence of three cases for $\sigma=0.2$ and $m=0.5, 1, 2$. The dashed line is for $\sigma=0.1$ and $m=1$; other cases for $m=0.5$ and 2 are very nearly the same.}
\end{figure}

\newpage
\begin{figure}[htbp]
\centering
\includegraphics[width=6in]{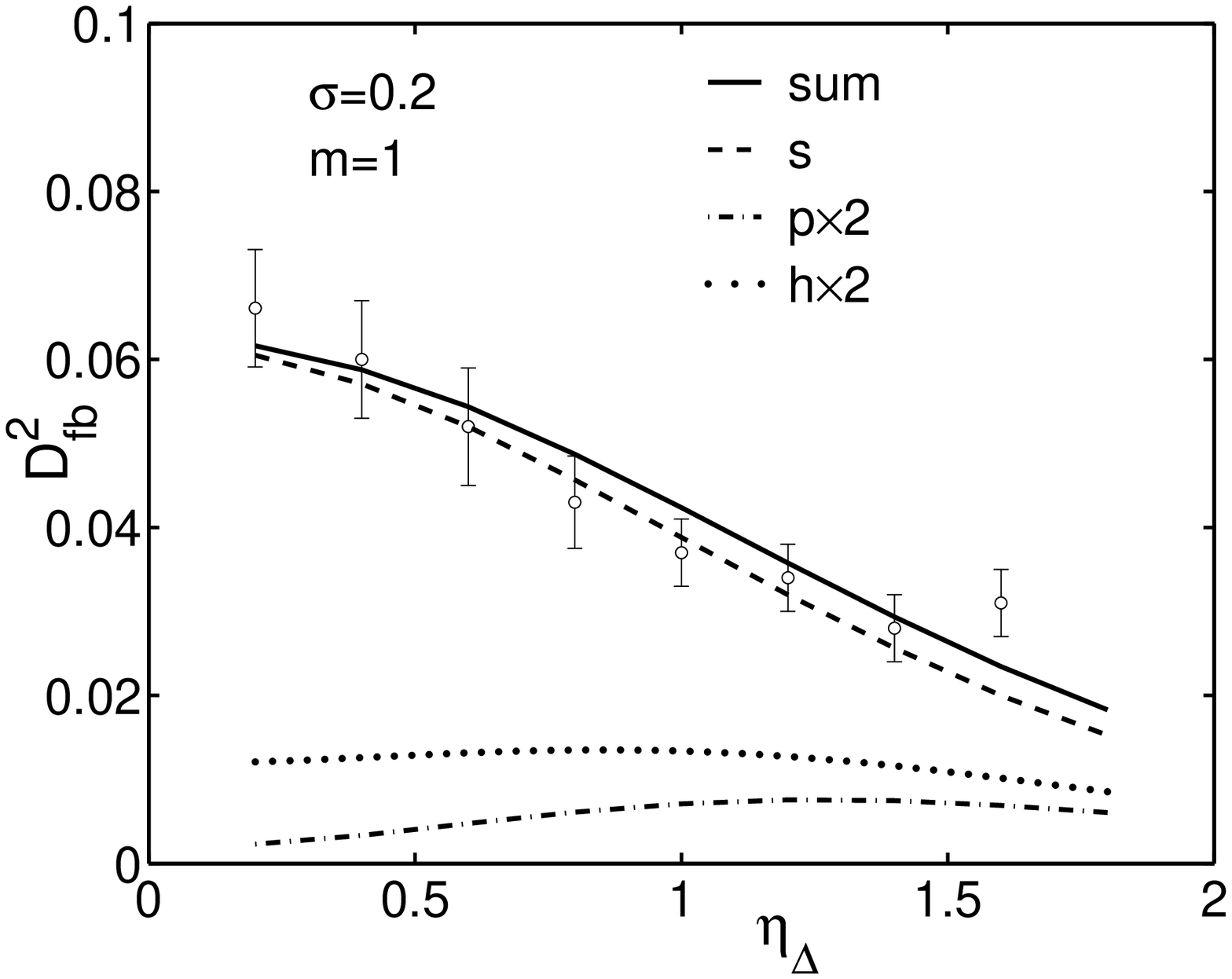}
\caption{Contributions to $D_{fb}^2$ from the $s$-wave component in dashed line and $p$-wave component in dashed-dotted line, raised by a factor of 2 for visual clarity. The solid line is their sum. The dotted line represents the (hard) component specified by $A_h$ in \eq{34}), also magnified by 2. The data are from \cite{sr}.}
\end{figure}


\newpage
\begin{thebibliography}{000}

\bibitem{kd}For a review see W.\ Kittel and E.\ A.\ De Wolf, {\it Soft Multihadron Dynamics} (World Scientific, Singapore, 2005).

\bibitem{cf}See the Proceedings of the Workshop on {\it Correlations and Fluctuations in Relativistic Nuclear Collisions}, Florence, Italy, July 2006, Proceedings of Science.

\bibitem{jd1}J.\ Adams et al. (STAR Collaboration), Phys.\ Lett.\ B {\bf 634}, 347 (2006).

\bibitem{jd2}J.\ Adams et al. (STAR Collaboration), Phys.\ Rev.\ C {\bf 73}, 064907 (2006).

\bibitem{jd3}J.\ Adams et al. (STAR Collaboration), nucl-ex/0605021.

\bibitem{tat} T.\ A.\ Trainor (STAR Collaboration), in Proc. of the 20th Winter Workshop on Nuclear Dynamics, Jamaica, 2004.

\bibitem{jd4}J.\ Adams et al. (STAR Collaboration), J.\ Phys. G {\bf 32}, L37 (2006).
 
\bibitem{pt} R.\ J.\ Porter and T.\ A.\ Trainor, J.\ Phys.\ Conf.\ Ser.\ {\bf 27}, 98 (2005).

\bibitem{pt1} R.\ J.\ Porter and T.\ A.\ Trainor, Acta Phys.\ Polonica B {\bf 36}, 353 (2005).

\bibitem{pt2} R.\ J.\ Porter and T.\ A.\ Trainor, in Ref.\ \cite{cf}, PoS (CFRNC 2006) 004.

\bibitem{tpp} T.\ A.\ Trainor, R.\ J.\ Porter and D.\ J.\ Prindle, J.\ Phys.\ G {\bf 31}, 809 (2005).

\bibitem{tp1}T.\ A.\ Trainor and D.\ J.\ Prindle, in Ref.\ \cite{cf}, PoS (CFRNC 2006) 009.

\bibitem{tp2}D.\ J.\ Prindle and T.\ A.\ Trainor, in Ref.\ \cite{cf}, PoS (CFRNC 2006) 007.

\bibitem{dpm}A.\ Capella, U.\ Sukhatme, C.-I.\ Tan, and J.\ Tran Thanh Van, Phys.\ Rep.\ {\bf 236}, 225 (1994).

\bibitem{lund}B.\ Andersson, G.\ Gustafson, G.\ Ingelman, T.\ Sj\"ostrand, Phys.\ Rep.\ {\bf 97}, 31 (1983).

\bibitem{aap}N.\ S.\ Amelin, N.\ Armesto, C.\ Pajares and D.\ Sousa, Eur.\ Phys.\ J.\ C {\bf 22}, 149 (2001).

\bibitem{aps}N.\ Armesto, C.\ Pajares and D.\ Sousa, Phys.\ Lett.\ B {\bf 527}, 92 (2002).

\bibitem{amp}N.\ Armesto, L.\ McLerran and C.\ Pajares, hep-ph/0607345.

\bibitem{ja5}J.\ Adams et al. (STAR Collaboration), Phys.\ Rev.\ D {\bf 74}, 032006 (2006).

\bibitem{sr} B.\ K.\ Srivastava (for STAR Collaboration), talk given at the International Workshop on Correlation and Fluctuation, Hangzhou, China 2006, nucl-ex/0702054.

\bibitem{rh}R.\ C.\ Hwa, Phys.\ Rev.\ D {\bf 22}, 1593 (1980).

\bibitem{hy}R.\ C.\ Hwa and C.\ B.\ Yang, Phys.\ Rev.\ C {\bf 66}, 025205 (2002); Phys.\ Rev.\ C {\bf 70}, 024905 (2004); Phys.\ Rev.\ C {\bf 73}, 044913 (2006).

\bibitem{rf}R.\ P.\ Feynman, Phys.\ Rev.\ Lett.\ {\bf 23}, 1415 (1969); in {\it High Energy Collisions}, edited by C.\ N.\ Yang et al., (Gordon and Breach, N.\ Y.\ 1969), p.\ 237.


\end{thebibliography}
\end{document}